\documentclass[useAMS,usenatbib,usegraphicx]{mn2e}

\newcommand{\dm}{\mathrm {dm}}
\newcommand{\br}{\mathrm {b}}

\newcommand{\tot}{\mathrm{tot}}
\newcommand{\Htwo}{H$_2$~}

\title[The formation and gas content of high redshift galaxies and
minihalos] {The formation and gas content of high redshift galaxies
and minihalos}

\author[S. Naoz and R. Barkana]{S. Naoz and R. Barkana$^{1}$
\thanks{E-mail: smadar@wise.tau.ac.il (SN); barkana@wise.tau.ac.il
(RB)}\\ $^{1}$School of Physics and Astronomy, The Raymond and Beverly
Sackler Faculty of Exact Sciences,\\ Tel Aviv University, Tel Aviv
69978, ISRAEL}

\begin{document}

\pagerange{\pageref{firstpage}--\pageref{lastpage}} \pubyear{2006}

\maketitle

\label{firstpage}

\begin{abstract}
We investigate the suppression of the baryon density fluctuations
compared to the dark matter in the linear regime. Previous
calculations predict that the suppression occurs up to a
characteristic mass scale of $\sim 10^6$~M$_\odot$, which suggests
that pressure has a central role in determining the properties of the
first luminous objects at early times. We show that the expected
characteristic mass scale is in fact substantially lower (by a factor
of $\sim 3$--10, depending on redshift), and thus the effect of
baryonic pressure on the formation of galaxies up to reionization is
only moderate. This result is due to the influence on perturbation
growth of the high pressure that prevailed in the period from cosmic
recombination to $z\sim 200$, when the gas began to cool adiabatically
and the pressure then dropped. At $z\sim10$ the suppression of the
baryon fluctuations is still sensitive to the history of pressure in
this high-redshift era. We calculate the fraction of the cosmic gas
that is in minihalos and find that it is substantially higher than
would be expected with the previously-estimated characteristic
mass. Expanding our investigation to the non-linear regime, we
calculate in detail the spherical collapse of high-redshift objects in
a $\Lambda$CDM universe. We include the gravitational contributions of
the baryons and radiation and the memory of their kinematic coupling
before recombination. We use our results to predict a more accurate
halo mass function as a function of redshift.
\end{abstract}

\begin{keywords}
galaxies:high-redshift -- cosmology:theory -- galaxies:formation
\end{keywords}

\section{Introduction}\label{intro}

The detection of the cosmic microwave background (CMB) temperature
anisotropies \citep{bennett} confirmed the notion that the present-day
galaxies and large-scale structure (LSS) evolved from the primordial
inhomogeneities in the density distribution at very early times.
After cosmic recombination, the gas decoupled from its mechanical drag
on the CMB, and the baryons subsequently began to fall into the
pre-existing gravitational potential wells of the dark matter. Regions
that were denser than average collapsed and formed bound halos. First
the smallest, least massive objects collapsed, and later, larger
objects formed through a mixture of mergers and accretion. The
formation and properties particularly of early galaxies at high
redshift are being actively studied in anticipation of many expected
observational probes \citep[e.g.,][]{BL01,R06}.

A well-known solution for the collapse of a halo that consists of dark
matter \emph{only} in an Einstein de Sitter (EdS) universe was
presented by \citet{gg}. This solution considers a spherical region
initially with a small uniform overdensity compared to the background
universe. As the universe expands, the overdensity expands slower than
the background until it reaches a maximum radius, turns around, and
collapses. The critical overdensity, in the corresponding
linearly-extrapolated calculation, marks the collapse time of a dark
matter halo in this case as $\delta_c=1.686$, a value that does not
depend on the halo mass or collapse redshift. The mathematical
solution gives a singularity as the final state, but physically we
know that even a small initial asymmetry will make the object
stabilize with a finite size after reaching a virial equilibrium
between motion and gravity. 

Extensive work has been done on spherical collapse models, especially
models that include a cosmological constant or a dark energy
background \citep[e.g.,][]{lahav,desh,hoffman,H,lahav2,wang}; in
particular, the cosmological constant $\Lambda$ changes the above
value of overdensity ($\delta_c$) by about $0.6\%$. In addition, many
numerical simulations of the formation of primordial objects at $z\sim
20$--30 have been performed. However, the earliest stars formed at $z
\sim 65$ \citep{NNB}, and even for a halo that collapsed at $z\sim30$,
$\delta$ must have started significantly non-linear ($\sim 9\%$) even
for a simulation that begins as early as $z \sim 600$.

With the $\Lambda$CDM cosmological parameters \citep{Spergel06}, the
contribution of the photons to the expansion of the universe cannot be
neglected when considering the formation of the first objects
\citep{NNB}. Moreover, the baryons have a non-negligible contribution
compared to the dark matter, and their different evolution must be
included in the collapse process.

When considering the formation and properties of the first luminous
objects, we must investigate the relation between the baryon and the
dark matter fluctuations. \citet{cs} defined a fiducial ``filtering
mass'' that describes the highest mass at which the baryonic pressure
still manages to suppress the linear baryonic fluctuations
significantly. \citet{gnedin00} extended the usefulness of the
filtering mass to the fully non-linear regime by showing that it is
also related to another characteristic mass scale -- the largest halo
mass for which the gas content is substantially suppressed compared to
the cosmic fraction. As we show below, if we follow previous
calculations \citep{cs,gnedin00,gnedin03}, we find a characteristic
mass at high redshift of $\sim 10^6 M_\odot$, approximately constant
at $z\ga 60$ and decreasing only slowly with time afterwards. This is
somewhat larger than the mass scale of the first objects and suggests
a potent effect on the formation of the first objects.

Here we present an improved calculation of the characteristic mass
that is mainly based on the improved calculation of the baryon density
and temperature fluctuations that we presented in \citet{NB}. We first
review the basic equations of linear perturbation growth
(Section~\ref{sec:scales}). We then divide the power spectrum into
several different ranges of scales that are associated with
large-scale structure (Section~\ref{LS}), the filtering scale
(Section~\ref{sec:TmTran}), and small scales
(Section~\ref{small}). Note that we define the filtering mass with a
different normalization than in previous works, as explained in
Section~\ref{sec:TmTran}. For completeness we compare our calculation
to the older, inaccurate approximation of a spatially-uniform sound
speed along with other approximations (Sections~\ref{sec:CS} and
\ref{sec:CSMF}). We use our results for the filtering mass to estimate
the gas fraction in minihalos (Section~\ref{sec:gas}). In
Section~\ref{non-linear} we calculate in detail the critical
overdensity for collapse of halos that form at very high redshifts,
following the evolution of perturbations outside the horizon
(Section~\ref{sec:outH}) and inside it (Section~\ref{sec:delc}). We
also predict the halo abundance at different redshifts
(Section~\ref{mass_a}). Finally, we summarize and discuss our results
in Section~4.

Our calculations are made in a $\Lambda$CDM universe, including dark
matter, baryons, radiation, and a cosmological constant. We assume
cosmological parameters matching the three year WMAP data together
with weak lensing observations \citep{Spergel06}, i.e.,
$\Omega_m=0.299$, $\Omega_\Lambda=0.74$, $\Omega_b=0.0478$, $h=0.687$,
$n=0.953$ and $\sigma_8=0.826$. We also consider the effect of current
uncertainties in the values of cosmological parameters on some of our
results, by comparing to the results with a different cosmological
parameter set specified by \citet{Viel}: $\Omega_m=0.253$,
$\Omega_\Lambda=0.747$, $\Omega_b=0.0425$, $h=0.723$, $n=0.957$ and
$\sigma_8=0.785$. These parameters represent typical 1-$\sigma$
errors, in terms of the parameter uncertainties given by
\citet{Spergel06}.

\section{Linear Growth of Perturbations}\label{sec:linear}

\subsection{The Basic Equations}\label{sec:scales}

\citet{NB} showed that the baryonic sound speed varies spatially, so
that the baryon temperature and density fluctuations must be tracked
separately. Thus, the evolution of the linear density fluctuations of
the dark matter ($\delta_\dm$) and the baryons ($\delta_\br$) is
described by two coupled second-order differential equations:
\begin{eqnarray}
\label{g_T}
\ddot{\delta}_{\dm} + 2H \dot {\delta}_{\dm} & = &
\frac{3}{2}H_0^2\frac{\Omega_{m}}{a^3}
\left(f_{\br} \delta_{\br} + f_{\dm} \delta_{\dm}\right)\ , \\
\ddot{\delta}_{\br}+ 2H \dot {\delta}_{\br} & = &
\frac{3}{2}H_0^2\frac{\Omega_{m}}{a^3} \left(f_{\br}
\delta_{\br} + f_{\dm}
\delta_{\dm}\right)-\frac{k^2}{a^2}\frac{k_B\bar{T}}{\mu}
\left(\delta_{\br}+\delta_{T}\right)\ ,\nonumber
\end{eqnarray}
where $\Omega_m$ is the present matter density as a fraction of the
critical density, $k$ is the comoving wavenumber, $a$ is the scale
factor, $\mu$ is the mean molecular weight, $H_0$ marks the present
value of the Hubble constant $H$, and $\bar{T}$ and $\delta_T$ are the
mean baryon temperature and its dimensionless fluctuation,
respectively. These equations can be derived by linearizing the
continuity, Euler, and Poisson equations. The baryon equation includes
a pressure term whose form comes from the equation of state of an
ideal gas. The linear evolution of the temperature fluctuations is
given by \citep{BL05,NB}
\begin{equation}
\label{gamma} \frac{d \delta_T} {d t} = \frac{2}{3} \frac{d
\delta_\br} {dt} + \frac{x_e(t)} {t_\gamma}a^{-4} \left\{
\delta_\gamma\left( \frac{\bar{T}_\gamma}{\bar{T}} -1\right)
+\frac{\bar{T}_\gamma} {\bar{T}} \left(\delta_{T_\gamma} -\delta_T
\right) \right\}\ ,
\end{equation}
where $x_e(t)$ is the electron fraction out of the total number
density of gas particles at time $t$, $\delta_\gamma$ is the photon
density fluctuation, $t_\gamma=8.55 \times 10^{-13}
{\mathrm{yr}}^{-1}$, and $T_\gamma$ and $\delta_{T_\gamma}$ are the
mean photon temperature and its dimensionless fluctuation,
respectively. Equation~(\ref{gamma}) results from the first law of
thermodynamics, where in the post-recombination era before the
formation of galaxies, the only external heating arises from Compton
scattering of the remaining free electrons with CMB photons. The first
term of equation~(\ref{gamma}) comes from the adiabatic cooling or
heating of the gas, while the second term is the result of the Compton
interaction. Note that prior analyses \citep[e.g.,][]{Peebles,Ma} had
assumed a spatially uniform speed of sound for the gas, but this
assumption is inaccurate.

It is often useful to express the evolution in terms of two linear
combinations of the dark matter and baryon fluctuations. Defining
$\delta_{\mathrm{tot}}=f_\br\delta_\br+f_\dm\delta_\dm$ (in terms of
the cosmic baryon and dark matter mass fractions $f_\br$ and $f_\dm$)
and $\Delta=\delta_\br-\delta_\mathrm{tot}$ \citep[following the
derivation in][]{BL05}, and using equations~(\ref{g_T}) we can write
two differential equations that describe the evolution of
$\delta_\mathrm{tot}$ and $\Delta$:
\begin{eqnarray}
\label{tot_ev_T}
\ddot{\delta}_{\mathrm{tot}} + 2H \dot {\delta}_{\mathrm{tot}}& =&
\frac{3}{2}H_0^2\frac{\Omega_{m}}{a^3}
\delta_\mathrm{tot}-f_\br\frac{k^2}{a^2}\frac{k_B\bar{T}}{\mu}
\left(\delta_\tot+\Delta+\delta_T\right)\ ,\nonumber \\
\ddot{\Delta}+2H\dot{\Delta}&=&-f_\dm\frac{k^2}{a^2}
\frac{k_B\bar{T}}{\mu}\left(\delta_\tot+\Delta+\delta_T\right)\ .
\end{eqnarray}
Note that the gravitational force depends directly on
$\delta_\mathrm{tot}$, which is the fluctuation in the total matter
density, while $\Delta$ describes the difference between the baryon
fluctuations and $\delta_\mathrm{tot}$. Before recombination, the
baryons are dynamically strongly coupled to the photons while the dark
matter fluctuations continue to grow independently. At lower redshifts
the dominant contribution to the baryon fluctuation growth is the
gravitational attraction to the dark matter gravitational wells (e.g.,
see Figure~1 in \citet{NB}).

\subsection{Large-Scale Structure: Small-$k$ Limit }\label{LS}

\begin{figure}
\includegraphics[width=84mm]{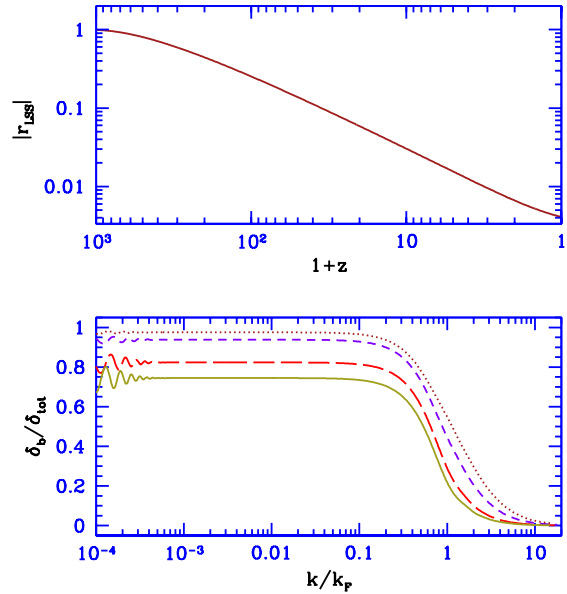}
\caption{The top panel shows $|r_{\rm LSS}|$ (equation~(\ref{r_LSS}))
versus redshift. The bottom panel shows the baryon to total
fluctuation ratio versus $k/k_F$ for various redshifts. We consider,
from bottom to top, $z=100$, 65, 20, and 7.}
\label{fig:db_tot}
\end{figure}

In the small-$k$ limit, the pressure terms can be neglected and
equations~(\ref{tot_ev_T}) can be approximated as
\citep[following the derivation in][]{BL05}
\begin{eqnarray}
\label{tot_ev2}
 \ddot{\delta}_{\mathrm{tot}} + 2H \dot {\delta}_{\mathrm{tot}}& =&
 \frac{3}{2}H_0^2\frac{\Omega_{m}}{a^3}\delta_\mathrm{tot}\ ,\\
 \nonumber \ddot{\Delta}+2H\dot{\Delta}&=&0\ .
\end{eqnarray}
Each of these equations has two independent solutions. Assuming that
the universe is accurately described as EdS before reionization, these
solutions are simple. The solutions for the $\delta_\mathrm{tot}$
equation are growing and decaying modes $D_{\mathrm{tot},1}\propto a$
and $D_{\mathrm{tot},2}\propto a^{-3/2}$, while for the $\Delta$
equation we obtain $D_{\Delta,1}\propto 1$ and $D_{\Delta,2}\propto
a^{-1/2}$. Moreover, with standard inflationary initial conditions,
equations~(\ref{tot_ev2}) satisfy $\Delta(k) \propto \delta_\tot(k)$
at a given redshift. In other words, the relation between $\Delta$ and
$\delta_\tot$ is independent of wavenumber, for the range of redshifts
and $k$ values that we are considering here. We therefore find it
useful to define
\begin{equation}
r_{\rm LSS} \equiv \frac{\Delta}{\delta_\tot}\ ,
\label{r_LSS}
\end{equation}
in terms of the solutions of equations~(\ref{tot_ev2}) in the
large-scale structure regime. The ratio $r_{\rm LSS}$ (which is
negative) is independent of $k$ in this regime, and its magnitude
decreases in time approximately $\propto 1/a$, since $\Delta$ is
roughly constant and $\delta_\tot$ is dominated by the growing mode
$\propto a$. Figure~\ref{fig:db_tot} (top panel) shows $|r_{\rm LSS}|$
as a function of redshift in the regime of large-scale
structure. Although $|r_{\rm LSS}|$ decreases as $z\to 0$, the initial
difference between the dark matter and the baryon fluctuations has a
large effect on the filtering mass even at $z<10$, as we show
below. The small-$k$ regime ($k \ll k_F$) can be seen in
Figure~\ref{fig:db_tot} (bottom panel), in terms of the filtering
wavenumber $k_F$ which is defined precisely in the following
subsection. Note that remnants of the acoustic oscillations in the
photon-baryon plasma can be seen on the largest scales ($k/k_F \sim
10^{-4}$, which corresponds to comoving $k \sim 0.01$--0.1
Mpc$^{-1}$), but these are much larger scales than those of
high-redshift halos and the oscillations do not affect our results.

A good numerical fit for $r_{\rm LSS}$ in the redshift range of
$z=7-150$ is given by
\begin{equation}
r_{\rm LSS}=\frac{\alpha_1(\Omega_m)}{a}+\frac{\alpha_2(\Omega_m)}{a^{3/2}}+
\alpha_3(\Omega_m)\ ,
\label{rLSS_fit}
\end{equation}
where the form is motivated by the definition of $r_{\rm LSS}$
(equation~(\ref{r_LSS})) and the EdS solutions of
equations~(\ref{tot_ev2}). The fitted coefficients are
\begin{eqnarray}\label{a_b_r_LSSfit}
\alpha_1(\Omega_m)&=&10^{-4}\times(-1.99\,\Omega_m^2+2.41\,
\Omega_m+0.21)\ ,\\ \nonumber
\alpha_2(\Omega_m)&=&10^{-3}\times(6.37\,\Omega_m^2-6.99\,
\Omega_m-1.76)\ ,\\ \nonumber
\alpha_3(\Omega_m)&=&10^{-2}\times(-1.83\,\Omega_m^2+2.4\,
\Omega_m-0.54)\ ,
\end{eqnarray}
where the maximum residual error of the fit is $0.2 \%$, over the
range $\Omega_m=$0.25--0.4 in $\Lambda$CDM.

\subsection{The Filtering Scale}\label{sec:TmTran}

In a $\Lambda$CDM universe, virialized CDM halos form on extremely
small scales at extremely early times. The minimum mass scale on which
galaxies form within these halos is determined through a combination
of two physical properties of the infalling gas. These are cooling and
pressure, and each produces a characteristic minimum scale, so that
the larger of the two scales becomes the dominant factor that
determines the minimum mass of star-forming halos.

During the gravitational collapse, the gas potential energy is
transformed into kinetic and thermal energy through virialization and
adiabatic compression. Unless the gas is able to dissipate its thermal
energy with an effective cooling mechanism, the temperature will
continue to rise, allowing the increasing pressure to halt the
collapse. At very high redshifts, before metals were produced in
supernovae and efficiently distributed in the intergalactic medium,
the only cooling mechanism at temperatures below 10,000 K was cooling
by molecular hydrogen, which itself is efficient only above a few
hundred K. Thus, the first objects are expected to be fairly
massive. Numerical calculations and simulations have shown that the
minimum cooling mass is $\sim 10^5 M_\odot$ at $z\sim 100$ and
increases with time \citep{th2,Abel,fuller,Yoshida}. An object more
massive than this minimum mass, after it goes through the
virialization process, will cool and collapse further, producing
high-density clumps in which stars can form.

\citet{Haiman} showed that the UV flux needed to dissociate \Htwo
in a collapsing environment is lower by more than two orders of
magnitude than the flux that is necessary to reionize the
universe. Therefore, once stars reach a certain abundance, the
formation of stars through \Htwo cooling is suppressed, and atomic
cooling becomes the only available mechanism. This mechanism requires
a much higher virial temperature ($T_{vir}\geq 10^4K$) which is
associated with a halo mass of $\sim10^8M_\odot$. Therefore, in order
to study the first generations of objects we must consider halo masses
of $\sim 10^5-10^8M_\odot$ (see also the review by \citet{R06}).

On large scales (small wavenumbers) gravity dominates halo formation
and pressure can be neglected. On small scales, on the other hand, the
pressure dominates, and the baryon density fluctuations are suppressed
compared to the dark matter fluctuations. The relative force balance at
a given time can be characterized by the Jeans scale, which is the
minimum scale on which a small perturbation will grow due to gravity
overcoming the pressure gradient. If the gas has a uniform sound speed
$c_s$, then the comoving Jeans wavenumber is
\begin{equation}
\label{eq:jeans}
k_J=\frac{a}{c_s}\sqrt{4\pi G\bar{\rho}}\ .
\end{equation}
Once the universe is matter-dominated, the Jeans scale is constant in
time as long as $T\sim1/a$, i.e., as long as the Compton scattering of
the CMB with the residual free electrons after cosmic recombination
keeps the gas temperature coupled to that of the CMB. At redshift
$z\sim 200$ the gas temperature decouples from the CMB temperature and
the Jeans scale decreases with time as the gas cools
adiabatically. Any halo more massive than the Jeans mass can begin to
collapse despite the pressure gradients. Figure~\ref{fig:kf} is a
reminder that the Jeans mass (dotted curve) is in the range $10^4$--
$10^5 M_\odot$ during the formation of the earliest generations of
galaxies.

The Jeans mass is related only to the evolution of perturbations at a
given time. When the Jeans mass itself varies with time, the overall
suppression of the growth of perturbations depends on a time-averaged
Jeans mass. Following \citet{cs}, we define a ``filtering'' scale and
use it to identify the largest scale on which the baryon fluctuations
are substantially suppressed compared to those of the dark
matter. While \citet{cs} assumed a spatially-uniform sound speed and
made a number of other approximations (see Section~\ref{sec:CS}), we
calculate here the fitering scale obtained from the exact numerical
solution of equations~(\ref{g_T}). In particular, our initial
conditions at high redshift account for the coupling of the baryons
and photons, i.e., $\delta_\br \ll \delta_\dm$ up to cosmic
recombination.

We define the filtering wavenumber $k_F$ based on the scale at which
the baryon-to-total fluctuation ratio drops substantially below its
value on large scales. Thus, we expand this ratio up to linear order
in $k^2$, and write the expansion in the following form:
\begin{equation}
\frac{\delta_\br}{\delta_\tot}=1-\frac{k^2}{k_F^2}+r_{\rm LSS}\ ,
\label{kf_btot}
\end{equation}
where $r_{\rm LSS}$ was defined in equation~(\ref{r_LSS}). Our
definition of $k_F$ is a generalization of that by \citet{cs}, who did
not include the $r_{\rm LSS}$ term. To find $k_F$ in our more general
case we first write it in the form:
\begin{equation}
k^2_F(t)=\frac{\delta_\tot(t)}{u(t)}\ , \label{kf}
\end{equation}
where $u(t)$ is to be determined. Substituting the expansion into
equations~(\ref{g_T}) and (\ref{tot_ev2}), we obtain an equation for
$u$:
\begin{equation}
\ddot{u}+2H\dot{u}=f_\dm\frac{1}{a^2}\frac{k_B\bar{T}}{\mu}
\left(\delta_\tot+r_{\rm LSS}\delta_\tot+\delta_T\right)\ ,
\label{ratio1}
\end{equation}
where we have neglected terms of higher order in $k^2$. We can solve
this equation to find the parameter $u$:
\begin{eqnarray}
\label{u_eq}
\lefteqn{
u(t)=\int^{t}_{t_{\rm rec}}\frac{dt^{\prime\prime}}{a^2(t^{\prime\prime})}
\int_{t_{\rm rec}}^{t^{\prime\prime}}
dt^{\prime}f_\dm\frac{k_B\bar{T}(t^\prime)}{\mu}} \\ \nonumber && \ \
\ \times\left(\delta_\tot(t^\prime)+r_{\rm LSS}(t^\prime)\delta_\tot
(t^\prime)+\delta_T(t^\prime)\right)\ .
\end{eqnarray}
We have started the integral from the time of cosmic recombination
($t_{\rm rec}$), since before recombination the contribution of the
baryon density and temperature fluctuations is negligible, i.e., the
integrand essentially vanishes (Note that $\delta_\tot+r_{\rm
LSS}\delta_\tot+\delta_T=\delta_\br+\delta_T$). As we discuss in
Section~\ref{sec:CSMF}, this integral at high redshift (before cosmic
reionization) gives a significantly different result from the
approximate formula of \citet{cs}.

Figure~\ref{fig:db_tot} (bottom panel) shows the baryon-to-total ratio
versus $k/k_F$ at several different redshifts. The different values in
the large-scale structure regime for different redshifts are due to
the different values of $r_{\rm LSS}$ (see top panel,
Figure~\ref{fig:db_tot}). The ratio drops on small scales. We have found
a functional form that can be used to produce a good fit for the drop
with wavenumber:
\begin{equation}
\frac{\delta_\br}{\delta_\tot}=(1+r_{\rm LSS})\left(1+\frac{1}{n}
\frac{x}{1+r_{\rm LSS}}\right)^{-n}\ ,
\end{equation}
where $x=k^2/k^2_F$, and $n$ must be adjusted at each
redshift. Defining
\begin{equation}
\label{def_y_eta}
\eta=\frac{1}{1+r_{\rm LSS}}\, \frac{k^2}{k_F^2}\ ;
\quad y=\frac{1}{1+r_{\rm LSS}}\, \frac{\delta_\br} {\delta_\tot}\ ,
\end{equation}
we can write the same fit as
\begin{equation}
\label{y_eta}
y=\left(1+\frac{1}{n}\eta\right)^{-n}\ .
\end{equation}
To second order, this gives: $y\approx 1-\eta+(n+1)\eta^2/(2n)$.

\begin{figure}
\includegraphics[width=84mm]{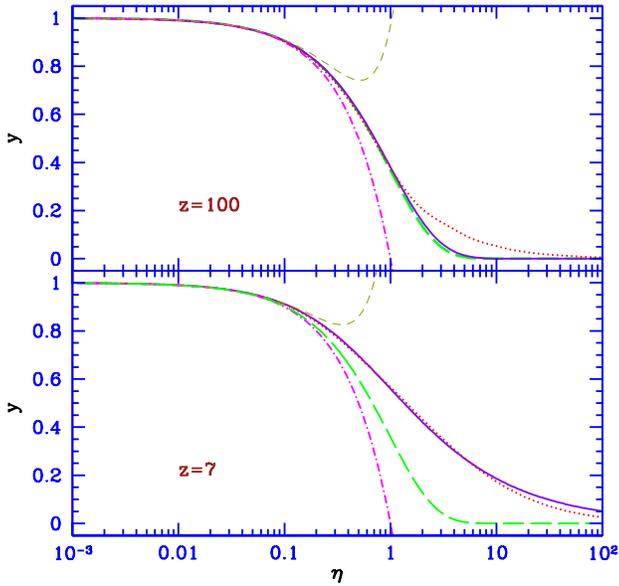}
\caption{$y$ versus $\eta$ for two redshifts, $z=100$ (top panel) and
$z=7$ (bottom panel). We compare the exact numerical results (dotted
curves) to several approximations: the formula of
equation~(\ref{y_eta}) (solid curves), the redshift-independent
first-order approximation (dot-dashed curves), the second-order
approximation (short-dashed curves), and the redshift-independent
exponential approximation (long-dashed curves) suggested by
\citet{gnedin03} for the post-reionization regime.}
\label{fig:y_eta}
\end{figure}

\begin{figure}
\includegraphics[width=84mm]{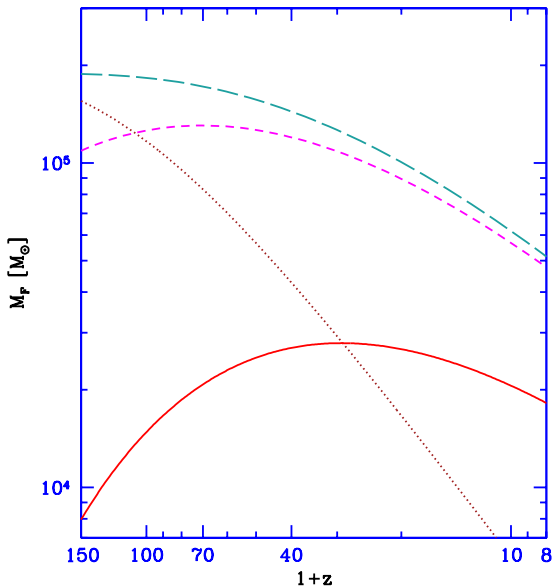}
\caption{The filtering mass versus redshift. The exact calculation
(solid curve), which can be fitted by equations~(\ref{Mf_fit}) and
(\ref{a_b_fit}), is compared to a couple versions of the mean sound
speed approximation (see Sections~\ref{sec:CS} and \ref{sec:CSMF}):
where the integrals in equation~(\ref{u_cs}) are begun at time zero as
written (long-dashed curve), or at cosmic recombination (short-dashed
curve). Also shown is the Jeans mass (dotted curve), which is constant
at $z\ga 150$.}
\label{fig:kf}
\end{figure}

We show in Figure~\ref{fig:y_eta} both the first and second-order
approximations, as well as the full formula of equation~(\ref{y_eta}),
compared to the exact numerical results.  We choose two redshifts that
bracket the interesting range, $z=100$ (bottom panel) for the dark
ages and $z=7$ (top panel) for the latest possible beginning of
reionization. From equation~(\ref{y_eta}) we can see that the
first-order approximation is independent of redshift. The figure also
shows the exponential approximation, generalized from the suggestion
made by \citet{gnedin03} for the post-reionization regime. In the
universe prior to reionization, the exponential approximation
$y=\exp(-\eta)$ (which is also independent of redshift) becomes highly
inaccurate at low redshifts. The figure makes clear that the drop of
the fluctuations with $k$ (or with $\eta$) has a functional form that
varies with redshift. With equation~(\ref{y_eta}) we find that the
power index at $z=100$ is $n=23$ (with behavior similar to an
exponential which would correspond to $n\to
\infty$), while at $z=7$ it is $n=0.5$. The values of $n$ were
found by matching the numerical result up to second order in
$\eta$.

We define the filtering mass in terms of the filtering wavenumber
using the traditional convention used to define the Jeans mass:
\begin{equation}
M_F=\frac{4\pi}{3}\bar{\rho_0}\left(\frac{1}{2}
\frac{2\pi}{k_F}\right)^3\ .
\label{Mf}
\end{equation}
Note that this relation, which we use consistently in this paper, is
one eighth of the definition in \citet{gnedin00}. The filtering mass
essentially describes the largest mass scale on which pressure must be
taken into account. Thus, in comparing different scenarios, those with
higher gas temperatures will tend to have higher pressures, leading to
higher values of the filtering mass.

Figure~\ref{fig:kf} shows the evolution of the filtering mass with
redshift. Since the baryon fluctuations are very small before cosmic
recombination, the gas pressure (which depends on $\delta_b$) is small
compared to gravity (which depends on $\delta_\tot$; see
equations~(\ref{g_T})). Thus, the filtering mass starts from low
values and rises with time at $z \sim 100$. At lower redshifts the gas
cools and the pressure drops. Therefore, even at $z \sim 10$ the
integral in equation~(\ref{u_eq}) receives a large contribution from
much higher redshifts ($z > 100$).

We have found a simple, accurate fit for the evolution of the
filtering mass in the redshift range of $z=7-150$. Using the notation
$L_M \equiv \log(M_F/M_\odot)$ and $L_z \equiv \log(1+z)$, the fit is
of the form
\begin{equation}
L_M=\beta_1(\Omega_m)L_z^3+\beta_2(\Omega_m)L_z^2+
\beta_3(\Omega_m)L_z+\beta_4(\Omega_m).
\label{Mf_fit}
\end{equation}
The fitted coefficients are
\begin{eqnarray}
\label{a_b_fit}
\beta_1(\Omega_m)&=&-0.38\, \Omega_m^2+0.41\, \Omega_m-0.16
\ ,\\ \nonumber
\beta_2(\Omega_m)&=&3.3\, \Omega_m^2-3.38\, \Omega_m+1.15
\ ,\\ \nonumber
\beta_3(\Omega_m)&=&-9.64\, \Omega_m^2+9.75\, \Omega_m-2.37
\ ,\\ \nonumber
\beta_4(\Omega_m)&=&9.8\, \Omega_m^2-10.68\, \Omega_m+11.6\ ,
\end{eqnarray}
where the maximum residual error of the fit is $0.2\%$, over the range
$\Omega_m=$0.25--0.4 in $\Lambda$CDM.

\subsection{Small Scale, Large-$k$ Limit}\label{small}

In this limit the pressure dominates and makes $\delta_b\ll
[\delta_\tot,\delta_\dm]$, and therefore the first
equation~(\ref{g_T}) becomes:
\begin{equation}
\label{largek_x}
 \ddot{\delta}_{\dm} + 2H \dot {\delta}_{\dm} \cong f_\dm
 \frac{3}{2}H_0^2\frac{\Omega_{m}}{a^3}\delta_\dm\ .
\end{equation}
Equation~(\ref{largek_x}) has a simple analytical solution in the EdS
regime: $\delta_\dm \propto t^\alpha$, where
$\alpha=\left(-1\pm\sqrt{1+24f_\dm}\right)/6$. In particular, the
growing mode of the dark matter is reduced compared to the usual
$\delta_\dm \propto t^{2/3}$ solution, since while the baryons
contribute to the cosmic expansion rate (in the Friedmann equation)
they do not, on small scales, contribute to perturbation growth. Thus,
since the dark matter fluctuations on large scales grow by a factor of
$\sim 300$ between recombination and $z=10$, then for $f_\dm = 0.84$
the total growth of the dark matter fluctuations is reduced on small
scales by a factor of $\sim 1.8$ relative to large scales. This linear
calculation is limited to redshifts $\ga 10$, since even without
reionization, by $z=10$ the filtering scale becomes non-linear.

\subsection{Mean Sound Speed Approximation}\label{sec:CS}

\citet{NB} showed that the presence of spatial fluctuations in the
sound speed modifies the calculation of perturbation growth
significantly. Nevertheless, for completeness and ease of comparison
with previous results we compare the above analysis to earlier,
approximate calculations. Thus, we proceed by applying a similar
derivation as in the previous sections. In this approximation of a
uniform sound speed, however, the evolution of the density
fluctuations is described by a different set of coupled second order
differential equations:
\begin{eqnarray}
\label{g_cs}
 \ddot{\delta}_{\dm}
 + 2H \dot {\delta}_{\dm} & = & \frac{3}{2}H_0^2\frac{\Omega_{m}}{a^3}
\left(f_{\br} \delta_{\br} + f_{\dm} \delta_{\dm}\right)\ , \\ 
\ddot{\delta}_{\br}+ 2H \dot {\delta}_{\br} & = &
\frac{3}{2}H_0^2\frac{\Omega_{m}}{a^3} \left(f_{\br} \delta_{\br} +
f_{\dm}
\delta_{\dm}\right)-\frac{k^2}{a^2}c^2_s\delta_{\br}\ ,\nonumber
\end{eqnarray}
where $c_s^2=dp/d\rho$ is assumed to be spatially uniform (i.e.,
independent of $k$). With this assumption, the temperature
fluctuations (as a function of $k$) are simply proportional at any
given time to the gas density fluctuations:
\begin{equation}
\frac{\delta_T}{\delta_{\br}}=\frac{c_s^2}{k_B \bar{T}/\mu}-1\ .
\end{equation}
This leads to different expressions for the filtering mass.

\subsection{Mean Sound Speed Approximation: The Filtering Scale}
\label{sec:CSMF}

We again expand the ratio of the baryon fluctuations to dark matter
fluctuations in powers of $k^2$. In this case we follow the derivation
in \citet{cs} and thus make several additional assumptions: that the
baryon fraction is small $(f_\br \ll f_\dm)$, that the dark matter
perturbation growth is dominated by the growing mode ($\delta_\dm
\propto D_+$), and that there is no initial difference between
the baryon and the dark matter fluctuations. Although this derivation
includes several different approximations, for simplicity we refer to
it as the ``mean sound speed approximation''. With these assumptions,
\begin{equation}
\frac{\delta_\br} {\delta_\dm}=1-\frac{k^2}{k_{F,c_s}^2}\ .
\label{kf_cs_bdm}
\end{equation}
Writing this as
\begin{equation}
\label{kf_cs}
k^2_{F,c_s}(t)=\frac{\delta_\dm(t)}{u_{c_s}(t)}\ ,
\end{equation}
and substituting into equations~(\ref{g_cs}), we obtain an equation
for the evolution of the parameter $u_{c_s}$ (analogous to
equation~(\ref{ratio1})):
\begin{equation}
\ddot{u}_{c_s}+2H\dot{u}_{c_s}=\frac{c_s^2}{a^2}D_+(t)\ .
\label{ratio1_cs}
\end{equation}
The solution (in analogy with equation~(\ref{u_eq})) is
\begin{equation}
\label{u_cs}
u_{c_s}(t)=\int^{t}_0\frac{dt^{\prime\prime}}{a^2(t^{\prime\prime})}\int
_0^{t^{\prime\prime}} dt^{\prime}c^2_s(t^\prime)D_+(t^\prime)\ .
\end{equation}
We note that the lower limit of the integral here is $z\to\infty$, and
not recombination as in equation~(\ref{u_eq}). Before recombination,
the coupling of the baryons with the radiation suppresses the baryon
density fluctuations, but this is unaccounted for in this approximate
calculation. Indeed, this formula implicitly assumes that the baryon
perturbations grow like those of the dark matter, except for the
effect of pressure. In particular, the filtering scale in this
approximation does not depend on the relative contributions of the
baryons and the dark matter to the total matter density.

In Figure~\ref{fig:kf}, we show the filtering mass with this
approximation from \citet{cs}. The correct filtering mass at $z \sim
10$ is substantially lower than predicted by the mean sound speed
approximation. As noted before, this is mainly due to the fact that
pressure forces depend on gradients in the gas density, and the baryon
perturbations are reduced on all scales until they catch up with the
dark matter fluctuations (at $z \ll 100$). In the approximate model,
on the other hand, the initial difference between the baryon and the
dark matter fluctuations is incorrectly neglected.

In the Figure, we consider also the mean sound speed approximation
where we start the integral in equation~(\ref{u_cs}) at recombination
($z=1200$). This should be more realistic than starting at time zero,
since the baryon density fluctuation before recombination were
negligible. This causes a rise in the filtering mass at $z \sim 100$,
as in the correct calculation (i.e., the solid curve). However, the
rise at high redshift is still much too fast, since this approximation
still assumes that the baryons catch up with the dark matter
fluctuations immediately after recombination.


To summarize, the Figure shows that the difference between the the
correct calculation and the approximate one persists with time. This
difference is remembered through the integrated effect of pressure,
since at low redshift the Jeans mass is lower and thus the pressure is
lower as well. Therefore, the high redshifts contribute most to the
integrated pressure, and thus even though the difference between the
baryon and the dark matter fluctuations declines with time (e.g., see
Figure~\ref{fig:db_tot} top panel), the system still remembers the
initial difference.

\subsection{Gas Fraction}\label{sec:gas}

One useful application of the filtering mass is to the estimation of
the fraction of gas inside halos. \citet[his equation~(8)]{gnedin00}
estimated the mean baryonic mass $M_g$ in halos of total mass $M_\tot$
using a formula fitted to his post-reionization simulation:
\begin{equation}
\bar{M}_g(M_\tot,t)\approx \frac{f_\br M_\tot}
{\left[1+(2^{1/3}-1)M_c(t)/M_\tot\right]^3}\ ,
\end{equation}
where $M_c(t)$ is a characteristic halo mass that corresponds to a gas
fraction of $50\%$ of the cosmic baryon fraction. It is natural to
expect a close relation between the characteristic halo mass and the
filtering mass, since the gas fraction in a collapsing halo reflects
the amount of gas that was able to accumulate in the central,
collapsing region, during the entire extended collapse process. In
particular, if the Jeans mass changes suddenly, this does not
immediately affect then-collapsing halos. A change of pressure
immediately begins to affect gas motions (through the
pressure-gradient force), but has only a gradual, time-integrated
effect on the overall amount of gas in a given region.

As mentioned before, our definition of the filtering mass
(equation~(\ref{Mf})) is one eighth of the previous definition of
\citet{gnedin00}. Thus the characteristic mass that matches the
\citet{gnedin00} simulations corresponds to about 8 times our
filtering mass. Assuming that this is true also prior to reionization,
we can evaluate the total gas fraction in halos as a function of
redshift as
\begin{equation}
F_g(z)=\int{f_{ST}(M_\tot,z)\,\frac{\bar{M}_g(M_\tot,t)}{M_\tot}\,dS}
\ , \end{equation}
where $S=\sigma^2(M,z)$ is the variance and $f_{ST}$ is the
\citet{Sheth} function for the fraction of mass associated with halos
of mass $M$ (explicitly given in equation~(\ref{sheth}), below).

\begin{figure}
\includegraphics[width=84mm]{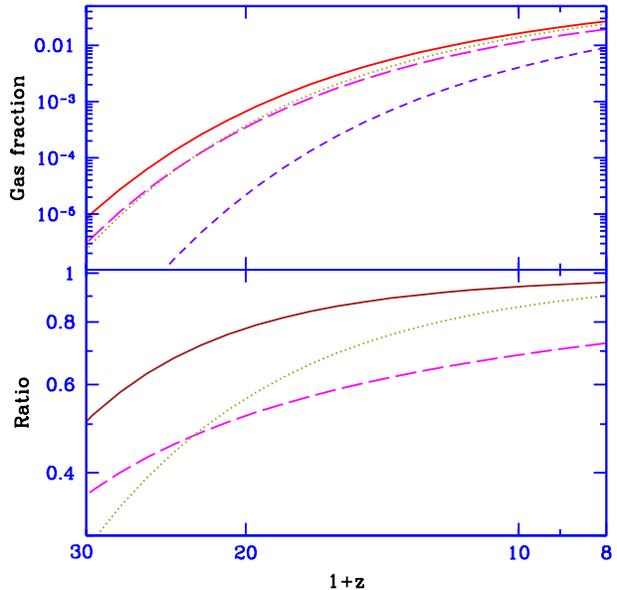}
\caption{Top panel: Gas fraction in halos versus redshift. We show 
our correct calculation (solid curve) and the previous calculation
using the filtering mass in the mean sound speed approximation (dotted
curve). We also compare to the fraction of gas in halos above the
$H_2$ cooling mass (long-dashed curve), and the fraction above the
atomic cooling mass (short-dashed curve). Bottom panel: Ratio of gas
fractions in different cases. We consider the gas fraction in halos
above the characteristic mass (solid curve), the gas fraction in halos
above the $H_2$ cooling mass (long-dashed curve), and the gas fraction
in the mean sound speed approximation (dotted curve); each is divided
by the total gas fraction in halos in our correct calculation.}
\label{fig:gas_frac}
\end{figure}

The total fraction of gas that is in halos is shown in
Figure~\ref{fig:gas_frac}, for our correct calculation of the filtering
mass as well as for the previous calculation (which we have referred
to as the mean sound speed approximation). We also compare to the
fraction of gas in halos above the minimum $H_2$ cooling mass, and to
the fraction above the minimum atomic cooling mass. In the correct
calculation, a significant part (10--$50\%$) of the total gas in halos
arises from halos that are below the characteristic mass. The
prediction of the gas fraction in halos in our correct calculation is
higher than that based on the previous approximation, by a factor $>2$
at high redshifts and still by $10\%$ at redshift 7. In this redshift
range, around half the gas in halos is in potentially star-forming
halos (i.e., those with efficient $H_2$ cooling), and the rest is in
gas minihalos.  In particular, this means that the smallest
star-forming halos (i.e., those with a mass equal to the minimum $H_2$
cooling mass) are moderately affected by pressure, and have their gas
content reduced to around half the cosmic baryon ratio. 

\begin{figure}
\includegraphics[width=84mm]{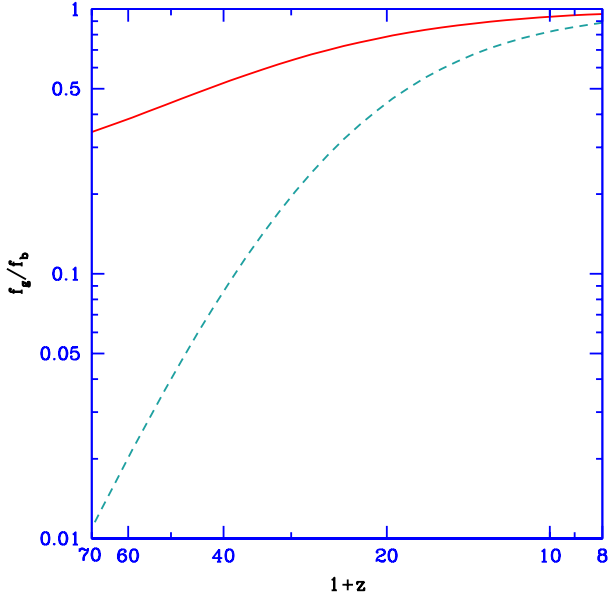}
\caption{The gas fraction in halos of mass equal to the minimum H$_2$
cooling mass compared to the cosmic fraction. We plot the improved
calculation (solid curve) and the mean sound speed approximation
(short-dashed curve). In the latter case we have used the previous
estimate, starting the integrals in eq.~(\ref{u_cs}) at $t=0$.}
\label{fig:fg_fb}
\end{figure}

The importance of the pressure in halos of mass equal to the minimum
H$_2$ cooling mass is illustrated in Figure~\ref{fig:fg_fb}. We
consider the improved calculation (solid curve) and the mean sound
speed approximation (short-dashed curve). This Figure shows that for
the very first stars the previous calculation underestimates the gas
fraction in these halos by more than an order of magnitude. E.g., at
$1+z=66$ (the formation of the first star) the improved calculation
predicts that the gas is about $35\%$ of the cosmic baryon fraction in
halos with mass equal to the H$_2$ cooling mass, while the mean sound
speed approximation predicts only $1.3\%$. Thus, we conclude that the
effect of pressure on the very first stars is only moderate, unlike
the result suggested by the mean sound speed approximation. The
discrepancy decreases with the redshift. However, at $1+z=20$ the
previous calculation still underestimates the gas fraction by a factor
of 2, and even at $1+z=8$ the previous calculation underestimates the
gas fraction by about $10\%$ compared with the improved calculation.

\section{Formation of Non-Linear Objects}\label{non-linear}

The small amplitude density fluctuations probed by the CMB grow over
time as described by equations~(\ref{g_T}) and (\ref{gamma}), as long
as the perturbations are linear. However, when $\delta$ in some region
becomes of order unity, the full non-linear problem must be
considered.
The standard calculation that describes the formation of spherical
non-linear objects was done for a dark matter halo \emph{only}, as was
explained in Section~\ref{intro}. Here, we generalize the calculation
to the high-redshift regime, including the gravitational effects of
the baryons and the radiation.

The initial conditions in N-body simulations of the first galaxies are
set long after the recombination epoch (usually at $z\sim200$ or
later). Halos that collapse early ($z\sim 20$ and earlier) thus cannot
arise from fluctuations that are linear at the beginning of the
simulation, regardless of the size of the corresponding perturbed
region. Now, perturbations that correspond to galaxies or clusters
start from a comoving length below 100 Mpc and
are thus expected to have entered the horizon at some time $t_{enter}$
in the radiation dominated era. The perturbations then do not grow
significantly until equality ($z\sim3000$). If we follow the evolution
of a collapsing halo, then the linear $\delta$ grows approximately
$\propto a$ after equality. Thus a region that collapses to a halo at
$z\sim100$ must have entered equality already in the weakly non-linear
regime ($\delta\sim 1.686/30 \sim 5\%$); since growth was slow before
then, even when the halo entered the horizon in the radiation
dominated universe, the perturbation was not extremely small. In
practical applications, we find that the largest $\delta$ gets at
horizon crossing is $\sim 10^{-3}$, for halos hosting the very first
stars \citep{NNB}, so that non-linear corrections are always small.
For completeness, however, we first consider the behavior of spherical
fluctuations outside the horizon.

\subsection{Fluctuation Growth Outside the Horizon}\label{sec:outH}

We consider a spherical top-hat fluctuation bigger than the Hubble
radius at an early cosmic time. Since the fluctuation is outside the
horizon, we cannot use Newtonian perturbation theory, and must apply
General Relativity.

We consider a spherical overdensity with a uniform density
$\rho=\bar{\rho}\left(1+\delta\right)$. 
Birkhoff's theorem implies that the Friedmann equation is the correct
solution within the spherical region (which has a curvature
$\tilde{k}$):
\begin{equation}
\label{gr}
H_{sph}^2 = H_0^2\left[
\frac{\Omega_m}{a^3}\left(1+\delta\right)+
\frac{\Omega_r}{a^4}\left(1+\delta\right)^\frac{4}{3} +
\frac{\Omega_{\tilde{k}}}{a_{sph}^2}+\Omega_\Lambda\right]\ ,
\end{equation}
where $H_{sph}$ and $a_{sph}$ are the Hubble constant and expansion
parameter, respectively, associated with the evolution of the
perturbed region, while $a$ describes the cosmic expansion of the mean
universe. The $4/3$ power is a result of the difference between the
non-relativistic (dark matter and baryon) density fluctuation $\delta$
and the radiation density fluctuation.

Just as in the regular top-hat collapse, we wish to compare the exact
non-linear evolution (given by equation~(\ref{gr})) to the
linearly-extrapolated evolution.  Comparing the above non-linear
equation to the evolution of the background universe and linearizing,
we obtain:
\begin{equation}
2H\alpha\dot{\delta}=I^2\delta-\frac{\tilde{k}}{a^2}, \label{gr_lin}
\end{equation}
where $I^2=H_0^2[\Omega_m/a^3+(4/3)\Omega_r/a^4]$, and
$\alpha=0.5\,I^2/\dot{H}$; at high redshift, in the
radiation-dominated regime, $\alpha\approx-0.25$ and then $\delta
\propto a^2$ (corresponding to the synchronous gauge; \citet{Ma};
\citet{pad}). We start early enough so that $\delta=1\times 10^{-4}$ 
initially, and assume the growing mode for the initial
perturbation. Solving numerically the non-linear and linear equations
(equations~(\ref{gr}) and (\ref{gr_lin})), until the fluctuation
enters the horizon, yields the initial value of the fluctuation
entering the horizon in the radiation dominated universe. As noted
above, in practice, linear initial conditions at horizon crossing
would suffice for the parameter space we consider in this paper.

\subsection{Fluctuation Growth Inside the Horizon}\label{sec:delc}

\begin{figure}
\includegraphics[width=84mm]{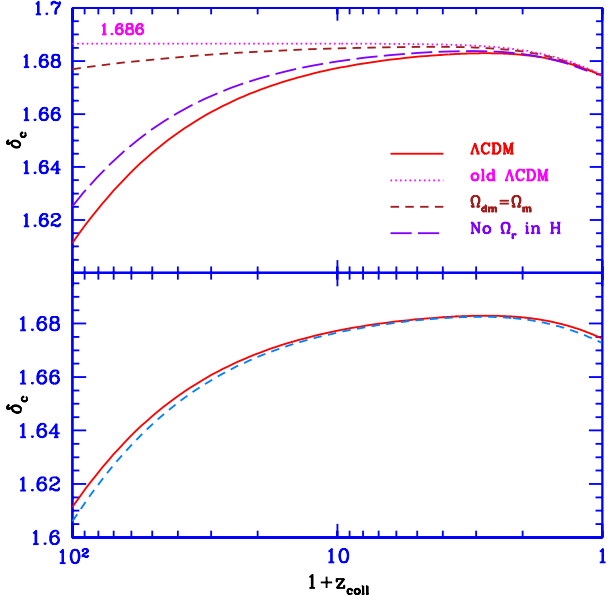}
\caption{Critical overdensity versus collapse redshift. We compare
$\delta_c$ for the full calculation (solid curves) to the results with
other assumptions. The top panel compares to several cases that
include only some of the physical ingredients that affect the
spherical collapse calculation. We show the results of previous
calculations that did not properly include the baryons and photons at
high redshift (dotted curve), the results if the baryons are treated
correctly but the contribution of the uniform radiation background is
neglected (long-dashed curve), and the results if the radiation is
included but the baryon-photon coupling is neglected (i.e., the
baryons are treated like dark matter: short-dashed curve). The bottom
panel shows results also for the \citet{Viel} set of parameters
(dashed curve).}
\label{fig:delta}
\end{figure}

In the Newtonian regime the non-linear growth is described by the
Newtonian equation (or, more precisely, the equation for the
acceleration that results from the Einstein equations):
\begin{equation}
\label{E2} \ddot{r}=-\frac{GM}{r^2}-\frac{4\pi
G}{3}\left(\rho+3P\right)_{rest}r\ ,
\end{equation}
where the {\em rest} stands for all matter that does not participate
in the collapse, and thus only contributes to the expansion of the
universe.  We define $r_\dm$ and $r_\br$ to be physical radii that
enclose a fixed mass of dark matter and of baryons, respectively,
assuming a tophat perturbation in each.
Then we obtain two coupled non-linear equations of motion:
\begin{eqnarray}
\label{non_r} \ddot{r}_\dm&=&\frac{-1}{r^2_\dm}\frac{4\pi G
}{3}r_\dm^3\left(\rho_\dm+\rho_\br\right)+H_0^2\Omega_\Lambda r_\dm-
\frac{8\pi G}{3}\rho_{r}r_\dm \ ,\nonumber \\
\ddot{r}_\br&=&\frac{-1}{r^2_\br}\frac{4\pi G
}{3}r_\br^3\left(\rho_\dm+\rho_\br\right)+H_0^2\Omega_\Lambda
r_\br-\frac{8\pi G}{3}\rho_{r} r_\br\ .
\end{eqnarray}
We have assumed that the radiation is kept smooth by its own pressure
and does not participate in the collapse; the factor $8\pi/3$ is the
result of inserting $P_r=\rho_r/3$ in equation~(\ref{E2}). Since the
time when the fluctuation enters the horizon is substantially early in
the radiation-dominated universe, the baryon-photon coupling yields
$\delta_\br,\delta_\gamma\ll\delta_\dm,\delta_\mathrm{tot}$
initially. We calculate separately the linear and non-linear growth of
the fluctuations. The resulting critical (linear) overdensity at the
time of collapse is shown in Figure~\ref{fig:delta}. For the
cosmological parameters that we use in this paper, we find that
$\delta_c$ is essentially independent of $M$, and is lower than the
EdS value by $\sim 1\%\times(1+z)/20$ in the range of $z=9-100$. When
dealing with very rare halos, even a change of a few percent in
$\delta_c$ can change the halo abundance at a given redshift by over
an order of magnitude (see Figure~2 of \citet{NNB}). The results are
insensitive to the cosmological parameters: also shown in
Figure~\ref{fig:delta} are the results for the set of other parameters
(Section~1) specified by \citet{Viel}, which differ by 1-$\sigma$ from
our standard WMAP parameters; the results are almost identical in the
EdS regime ($1+z=2-10$), with only a $\la 0.2\%$ difference at high
redshift (and even less at $z<1$).

The effect of various physical ingredients on $\delta_c$ can be
illustrated using the different cases shown in the upper panel of
Figure~\ref{fig:delta}. In general, collapse is most efficient in the
EdS case where all the matter participates in the collapse (resulting
in $\delta_c=1.686$); any smooth component that does not collapse (a
cosmological constant at low redshift, radiation and baryons at high
redshift) reduces the collapse efficiency since only the dark matter
component takes part in the collapse throughout. Now, any reduction in
the matter fraction that collapses depresses the linear evolution of
the density perturbation more strongly, while the non-linear
perturbation is larger and is thus less affected by the components
that do not help in the collapse. Therefore the linear perturbation
reaches a lower value of $\delta_c$ when the non-linear perturbation
collapses. As shown in the upper panel of Figure~\ref{fig:delta}, the
extended period at high redshift when the baryon perturbations remain
suppressed is the main cause of the reduction of the value of
$\delta_c$, but the contribution of the photons to the expansion of
the universe also makes a significant contribution.

\subsection{The Mass Function and Halo Abundance}\label{mass_a}

In addition to characterizing the properties of individual halos, it
is important for any model of structure formation to predict also the
abundance of halos. We calculate in this section the number density of
halos as a function of mass, at any redshift. We start with the simple
analytical model of \citet{Press}, which is based on Gaussian random
fields, linear growth and spherical collapse. It predicts that the
abundance of halos depends on mass and redshift through the two
functions $\sigma(M,z)$ and $\delta_c(z)$, where $\sigma^2(M,z)$ is
the variance (calculated from the power spectrum) as a function of
halo mass at $z$, and $\delta_c(z)$ is the critical collapse
overdensity from Section~\ref{sec:delc}.

\begin{figure}
\includegraphics[width=84mm]{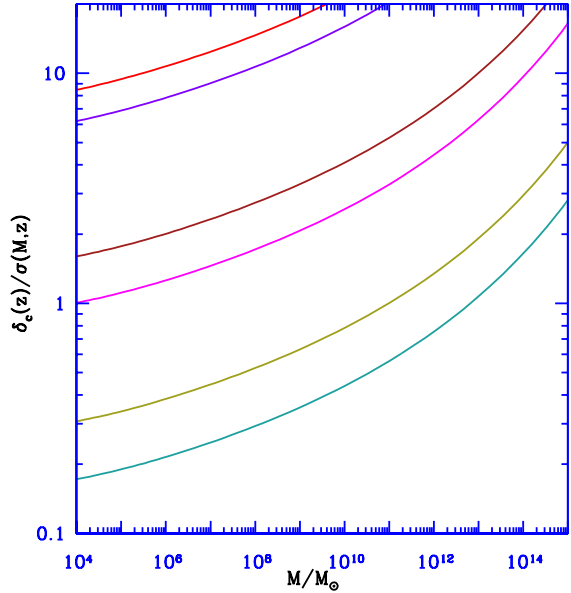}
\caption{Rarity of fluctuations that produce halos at various redshifts.
We show the size of a fluctuation that produces a collapsed halo on
the scale $M$ at redshift $z$, in terms of the typical fluctuation
level (i.e., measured as a number of $\sigma$). This includes
$\sigma(M,z)$ from the correct calculation of the power spectrum
\citep{NB}, and $\delta_c(z)$ from the correct spherical collapse
calculation (Section~\ref{sec:delc}). We consider, from bottom to top,
$z=0$, 1.2, 6.5, 11, 47 and 66. Redshift 66 corresponds to the
formation of the first star, while $z=47$ corresponds to the redshift
of the second generation of stars, i.e., the first collapse via atomic
cooling. Redshift 11 corresponds to the first halo as massive as that
of the Milky Way and $z=1.2$ is associated with the formation of the
first cluster as massive as Coma (see \citet{NNB}). We also show the
results at $z=6.5$ associated with observations of the most distant
quasars (and perhaps with the end of reionization).}
\label{fig:sigma_dc}
\end{figure}

In Figure~\ref{fig:sigma_dc} we show the number of $\sigma$ that a
fluctuation must be in order for it to collapse, at various redshifts
from the first star until the present. For example, a $10^5 M_\odot$
halo that collapses and forms the first star is a 9.5-$\sigma$
fluctuation according to this criterion (but see below). Halos of mass
$10^5 M_\odot$ become 1-$\sigma$ fluctuations only at $z\sim5$, while
today the 1-$\sigma$ collapsing fluctuations are $10^{13}M_\odot$
group halos.

\begin{figure}
\includegraphics[width=84mm]{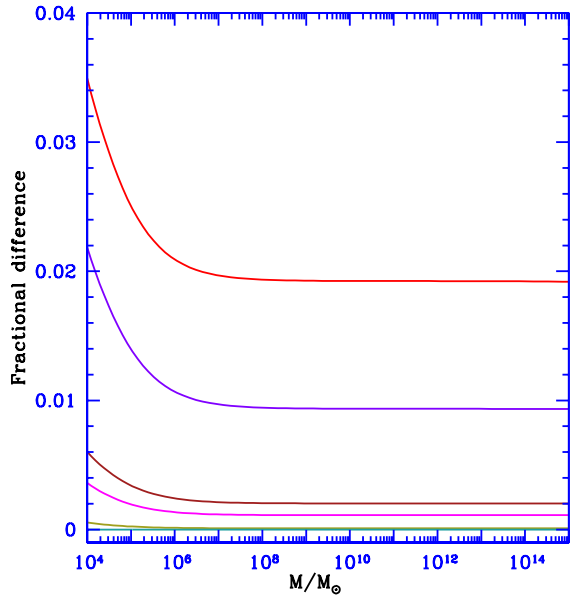}
\caption{The fractional difference between the
effective growth factor of our correct calculation,
$\sigma(M,z)/\sigma(M,z=0)$, and the traditional growth factor $D(z)$,
as a function of mass at various redshifts. We consider, from bottom
to top, $z=0$, 1.2, 6.5, 11, 47 and 66, as in
Figure~\ref{fig:sigma_dc}.}
\label{fig:S_D}
\end{figure}

The growth of the density fluctuations in our correct calculation can
be described by an effective growth factor which depends on the mass
scale: $D_{\rm eff} \equiv \sigma(M,z)/\sigma(M,z=0)$. We compare this
effective growth factor to the traditional growing mode $D(z)$
\citep{Peebles} in Figure~\ref{fig:S_D}, where we show
$D_{\rm eff}/D-1$ as a function of mass, illustrated at the same
redshifts as in Figure~\ref{fig:sigma_dc}. The source of the difference
is in the smoothness of the radiation and baryons at high redshifts;
in addition, pressure suppresses growth on small mass scales. These
effects make $D_{\rm eff}$ larger at high redshift compared to its
value today (since there has been less growth).

The comoving number density of halos of mass $M$ at redshift $z$
is
\begin{equation}
\label{dndM}
\frac{dn}{dM}=\frac{\rho_0}{M}f_{ST}\left|\frac{dS}{dM}\right|\ ,
\end{equation}
were we have used the \citet{Sheth} mass function that fits
simulations, and includes non-spherical effects on the collapse. The
function $f_{ST}$ is the fraction of mass associated with halos of
mass $M$:
\begin{equation}
\label{sheth}
f_{ST}(\delta_c,S)=A'\frac{\nu}{S}\sqrt{\frac{a'}{2\pi}}
\left[1+\frac{1}{\left(a'\nu^2\right)^{q'}} \right]
\exp \left[\frac{-a'\nu^2}{2} \right]\ ,
\end{equation}
where $\nu = \delta_c/\sqrt{S}$. We use best-fit parameters $a'=
0.75$ and $q'= 0.3$ \citep{Sheth02}, and ensure normalization to
unity by taking $A'= 0.322$. We apply this formula with
$\delta_c(z)$ and $\sigma^2(M,z)$ as the arguments. The
\citet{Sheth} mass function makes it easier for rare fluctuations
to collapse compared to \citet{Press}. This, for instance, makes the
first star-forming halo effectively only an 8.3-$\sigma$ density
fluctuation on the mass scale of $10^5 M_{\odot}$, in terms of the
total cosmic mass fraction contained in halos above this mass.

\begin{figure}
\centering
\includegraphics[width=84mm]{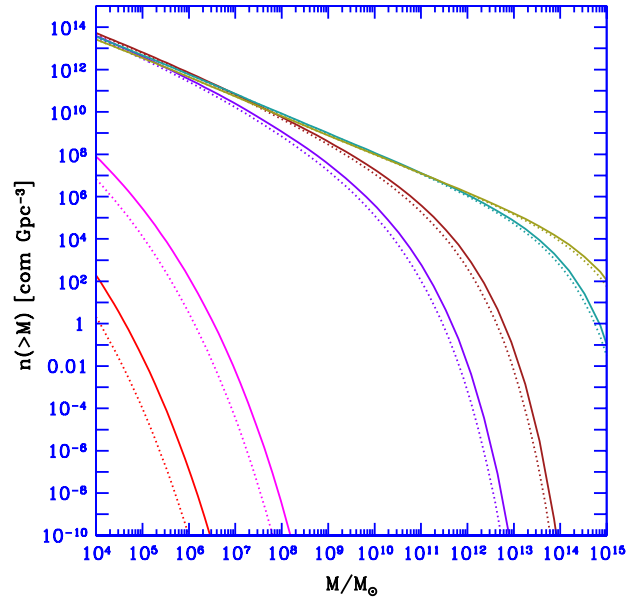}
\caption{The halo abundance as a function of the halo
mass at different redshifts. From top to bottom (at the high-mass
end): $z=0,1.2,6.5,11,47$ and finally $z=66$ which corresponds to the
formation of the first star \citep{NNB}. We consider our standard WMAP
cosmological parameters (solid curves), and compare to the
\citet{Viel} set of parameters (dotted curves).}
\label{fig:NofM}
\end{figure}

The cumulative comoving number density of halos $n(>M_{\rm min})$ is
given by
\begin{equation}
\label{nofM}
n(>M_{\rm min})=\int_{M_{\rm min}}^\infty \frac{dn}{dM}\,dM\ .
\end{equation}
In Figure~\ref{fig:NofM} we plot the cumulative number density of
halos. In addition to our standard WMAP cosmological parameter set, we
have also compared to the \citet{Viel} set of parameters
(Section~1). We obtained for these parameters that the first star
formed at $z=60.0$ (instead of 65.8) and the first Coma size cluster
formed at $z=1.06$ (instead of 1.24). The difference between the two
sets of cosmological parameter arises mainly from the difference in
$\sigma_8$. The difference in the values of $\delta_c$ is negligible
(see Figure~\ref{fig:delta}).

\section{Conclusions}

We have calculated the filtering mass correctly at high redshift and
compared it to previous estimates. We have found that at high redshift
the filtering mass is lower by about an order of magnitude compared to
previous calculations. The difference declines with time but remains a
factor of $\sim 3$ even at $z=7$ (Figure~\ref{fig:kf}). Our
calculation predicts a lower filtering mass because it includes the
initial difference between the dark matter and baryon fluctuations,
due to the baryon-photon coupling, and this lowers the pressure. This
means that in contrast with the previous prediction, pressure has only
a moderate effect on the formation of the earliest luminous objects.

Before recombination the baryon fluctuations were suppressed compared
to the dark matter fluctuations due to tight coupling with the
radiation. The filtering mass rises with time at high redshift because
after recombination the baryon pressure gradients start to
increase. However, after the baryon temperature decouples from the CMB
temperature the gas cools adiabatically, the Jeans mass drops, and
eventually (at lower redshifts) the filtering mass drops as well (see
Figure~\ref{fig:kf}). The high pressure at very high redshift still
contributes significantly to the filtering mass at redshift below
10. The delay in the drop of the filtering mass compared to the Jeans
mass is a signature of the continuing contribution of the memory of
the pressure from very high redshifts. We have found numerical fits to
the difference between the baryon and the total density fluctuations
on large scales (equations~(\ref{rLSS_fit})); we have also fit the
filtering mass as a function of $\Omega_m$ and $z$
(equations~(\ref{Mf_fit}) and (\ref{a_b_fit})).

Using the new prediction of the filtering mass we have shown that at
high redshift there is more gas in halos than in previous
estimates. This difference is as large as a factor of 2, but remains
$\sim 10\%$ even at redshift 7. Around half of the gas is in halos
with efficient $H_2$ cooling, and the rest is in gas minihalos. The
previous calculation also suggests a greatly reduced gas fraction in
the halo that hosts the first star, while we find only a moderate
effect (Figure~\ref{fig:fg_fb}).

In addition, we have computed the evolution of linear as well as
non-linear spherical overdensities, outside and inside the horizon, in
a $\Lambda$CDM universe. Previous analyses showed that the
cosmological constant contribution to the expansion of the universe
results in a drop of the value of $\delta_c$ by $\sim 0.6 \%$ today
\citep[e.g.,][]{lahav}. This makes a significant difference in the
abundance of clusters. Considering structure formation at high
redshift, we investigated the effect on $\delta_c$ and on the halo
abundance of the contribution of radiation to the expansion of the
universe, and of the contribution of the baryons to the collapsing
halo, given their different initial conditions compared to the dark
matter. We have found that there is a $3\%$ change in the value of the
overdensity $\delta_c$ at $z \sim 60$ (Figure~\ref{fig:delta}). This
changed value translates to a much larger difference in the halo
abundance at high redshift (Figure~\ref{fig:NofM}); these differences
decline at low redshift. The large difference in the halo abundance is
the result of the difference in the exponent in the mass function
(equation~(\ref{sheth})), and shows that even small effects on halo
formation can in some cases be very important.

\section*{Acknowledgments}

The authors acknowledge support by Israel Science Foundation grant
629/05 and U.S. - Israel Binational Science Foundation grant 2004386.

\bsp

\label{lastpage}

\end{document}